\documentclass{article}

\begin{document}
\begin{center}

NEGATIVE ENERGIES AND TIME REVERSAL

 IN QUANTUM FIELD THEORY
\vspace{1cm}

Frederic Henry-Couannier

CPPM, 163 Avenue De Luminy, Marseille 13009 France.

henry@cppm.in2p3.fr 
\end{center}


\begin{abstract}
The theoretical and phenomenological status of negative energies is reviewed in
 Quantum Field Theory leading to the conclusion that hopefully their rehabilitation might only be completed in 
a modified general relativistic model. 
\end{abstract}
\section{ Introduction}

With recent cosmological observations related to supernovae, CMB and 
galactic clustering the evidence is growing that our universe is undergoing 
an accelerated expansion at present. Though the most popular way to account 
for this unexpected result has been the reintroduction of a cosmological 
constant or a new kind of dark matter with negative pressure, scalar fields 
with negative kinetic energy, so-called phantom fields, have recently been 
proposed \cite{Cald} \cite{Fram} \cite{Carr} as new sources leading to the not excluded possibility that the 
equation of state parameter be less than minus one. Because such models 
unavoidably lead to violation of positive energy conditions, catastrophic 
quantum instability of the vacuum is expected and one has to impose an 
ultraviolet cutoff to the low energy effective theory in order to keep the 
instability at unobservable rate. Stability is clearly the challenge for any 
model trying to incorporate negative energy fields interacting with positive 
energy fields. But before addressing this crucial issue, it is worth 
recalling and analyzing how and why Quantum Field Theory discarded negative 
energy states. We shall find that this was achieved through several not so 
obvious mathematical choices, often in close relation with the well known 
pathologies of the theory, vacuum and UV loop divergences. Following another 
approach starting from the orthogonal alternative mathematical choices, the 
crucial link between negative energies, time reversal and the existence of 
discrete symmetry conjugated worlds will appear. 

\section{ Negative energy and classical fields}
\subsection{ Extremum action principle}

Let us first address the stability of paths issue. Consider the path r(t) of 
a material point of mass m with fixed endpoints at time t$_{1}$ and t$_{2}$ 
in the potential U(r,t). The action S is:
\[
S=\int_{t_1 }^{t_2 } {(1/2\,mv^2\;-U(r,t))dt} 
\]
The extremum condition~($\delta $S=0) is all we need to establish the 
equation of motion:
\[
m\dot {v}=-\frac{\partial U}{\partial r}
\]
S has no maximum because of the kinetic term positive sign.~The extremum we 
find is a minimum. 
Let us try now a negative kinetic term:
\[
S=\int_{t_1 }^{t_2 } {(-1/2\,m{\kern 1pt}\,v^2\;-U(r,t))dt} 
\]
The extremum condition~($\delta $S=0) is all we need to establish the 
equation of motion:

\begin{center}
-$m\dot {v}=-\frac{\partial U}{\partial r}$
\end{center}
S has no minimum because of the kinetic term negative sign.~The extremum we 
find is a maximum. 
Eventually, it appears that the fundamental principle is that of stationary 
($\delta $S=0) action, the extremum being a minimum or a maximum depending 
on the sign of the kinetic term. In all cases we find stable trajectories.

\subsection{ Classical relativistic fields}

We can also check that negative kinetic energy terms (ghost terms) in a free 
field action are not problematic. When we impose the extremum action 
condition the negative energy field solutions simply maximize the action. 
Now, in special relativity for a massive or mass-less 
particle, two energy solutions are always possible:
\[
E=\pm \sqrt {p^2+m^2} ,\,E=\pm \left| p \right|
\]
In other words, the Lorentz group admits, among others, negative energy 
representations $E^2-p^2=m^2>0,\;E<0$\textbf{,} $E^2-p^2=0,\;E<0$. 
Thus, not only can we state that negative energy free field terms are not 
problematic but also that negative energy field solutions are expected in 
any relativistic field theory. For instance the Klein-Gordon equation:
\[
\left( {\partial ^\mu \partial _\mu +m^2} \right)\mathop \phi \limits^{(\sim 
)} (x)=0
\]
admits when $m^2>0$ (we shall not try to understand here the physical 
meaning of tachyonic ($m^2<0$) and vacuum ($E=p=m=0$) representations) 
positive $\mathop \phi \limits (x)$ and negative $\mathop \phi \limits^\sim 
(x)$ energy free field solutions. Indeed, the same Klein-Gordon equation 
results from applying the extreme action principle to either the `positive' 
scalar action:
\[
\int {d^4x\,} \phi (x)\left( {\partial ^\mu \partial _\mu 
+m^2} \right)\phi (x)
\]
or the `negative' scalar action:
\[
-\int {d^4x\,} \tilde {\phi } (x)\left( {\partial ^\mu 
\partial _\mu +m^2} \right)\tilde {\phi }(x)
\]
From the former a positive conserved Hamiltonian is derived through the 
Noether theorem:
\[
\int {d^3x} (\frac{\partial \phi^\dag  ( \textbf{x} ,t)}{\partial 
t}\frac{\partial \phi (\textbf{x} ,t)}{\partial t}+\sum\limits_{i=1,3} 
{\frac{\partial \phi^\dag (\textbf{x} ,t)}{\partial x_i 
}\frac{\partial \phi (\textbf{x} ,t)}{\partial x_i }+m^2\phi 
^\dag (\textbf{x} ,t)\phi (\textbf{x} ,t)})
\]
while a negative one is derived from the latter:
\[
-\int {d^3x}(\frac{\partial \tilde {\phi^\dag } ( \textbf{x} ,t)}{\partial 
t}\frac{\partial \tilde {\phi }(\textbf{x} ,t)}{\partial t}+\sum\limits_{i=1,3} 
{\frac{\partial \tilde {\phi^\dag }(\textbf{x} ,t)}{\partial x_i 
}\frac{\partial \tilde {\phi }(\textbf{x} ,t)}{\partial x_i }+m^2\tilde {\phi 
}^\dag (\textbf{x} ,t)\tilde {\phi }(\textbf{x} ,t)})
\]

\section{Negative energy in relativistic Quantum Field Theory (QFT)}

\subsection{Creating and annihilating negative energy quanta}

At first sight it would seem that the negative frequency terms appearing in 
the plane wave Fourier decomposition of any field naturally stand for the 
negative energy solutions. But as soon as we decide to work in a 
self-consistent quantization theoretical framework, that is the second 
quantization one, the actual meaning of these negative frequency terms is 
clarified. Operator solutions of field equations in conventional QFT read:
\[
\phi (x)=\phi _+ (x)+\phi _- (x)
\]
with $\phi _+ (x)$ a positive frequency term creating \textbf{positive} 
energy quanta and $\phi _- (x){\rm }$ a negative frequency term annihilating 
\textbf{positive} energy quanta. So negative energy states are completely 
avoided thanks to the mathematical choice of creating and annihilating only 
positive energy quanta and $\phi (x)$ built in this way is just the positive 
energy solution. This choice would be mathematically justified if one could 
argue that there are strong reasons to discard the `negative action' we 
introduced in the previous section. But there are none and as we already 
noticed the Klein-Gordon equation is also easily derived from such action 
and the negative energy field solution:
\[
\tilde {\phi }(x)=\tilde {\phi }_+ (x)+\tilde {\phi }_- (x)
\]
(with $\tilde {\phi }_+ (x)$ a positive frequency term annihilating 
\textbf{negative} energy quanta and $\tilde {\phi }_- (x)$ a negative 
frequency term creating \textbf{negative} energy quanta) is only coherent 
with the negative Hamiltonian derived from the negative action through the 
Noether theorem (in the same way it is a standard QFT result that the usual 
positive energy quantum field $\mathop \phi \limits (x)$ is only coherent 
with the above positive Hamiltonian \cite{Wein2} \cite{Grei}). Therefore, it is mathematically 
unjustified to discard the negative energy solutions. Neglecting them on the 
basis that negative energy states remain up to now undetected is also very 
dangerous if we recall that antiparticles predicted by the Dirac equation 
were considered unphysical before they were eventually observed. If negative 
(or tachyonic) energy states are given a profound role to play in physics, 
this must be fully understood otherwise we might be faced with 
insurmountable difficulties at some later stage. 

There is a widespread belief that the negative energy issue were once and 
for all understood in terms of antiparticles. Indeed, because charged fields 
are required not to mix operators with different charges, the charge 
conjugated creation and annihilation operators (antiparticles) necessarily 
enter into the game. Following Feynman's picture, such antiparticles can as 
well be considered as negative energy particles propagating backward in 
time. According S.Weinberg \cite{Wein1}, it is only in relativistic (Lorentz 
transformation do not leave invariant the order of events separated by 
space-like intervals) quantum mechanics (non negligible probability for a 
particle to get from x$_{1}$ to x$_{2 }$ even if x$_{1}$ - x$_{2 }$ is 
space-like) that antiparticles are a necessity to avoid the logical paradox 
of a particle being absorbed before it is emitted. However, these 
antiparticles have nothing to do with genuine negative energy states 
propagating forward in time, whose quanta are by construction of the 
conventional QFT fields never created nor annihilated. Therefore, our deep 
understanding of the actual meaning of field negative frequency terms in QFT 
does not ``solve'' the negative energy issue since the corresponding 
solutions were actually neglected from the beginning. As we shall see, there 
is a heavy price to pay for having neglected the negative energy solutions: 
all those field vacuum divergences that unavoidably arise after quantization 
and may be an even heavier price are the ideas developed to cancel such 
infinities without reintroducing negative energy states. 

\subsection{A unitary time reversal operator}
In a classical relativistic framework, one could not avoid energy reversal 
under time reversal simply because energy is the time component of a 
four-vector. But, when one comes to establish in Quantum Field Theory the 
effect of time reversal on various fields, nobody wants to take this simple 
picture serious anymore mainly because of the unwanted negative energy 
spectrum it would unavoidably bring into the theory. It is argued that 
negative energy states remain undetected and that their existence would 
necessarily trigger catastrophic decays of particles and vacuum: matter 
could not be stable. To keep energies positive, the mathematical choice of 
an anti-unitary time reversal operator comes to the rescue leading to the 
idea that the time-mirrored system corresponds to `running the movie 
backwards' interchanging the roles of initial and final configurations. We 
shall come back to the stability issue later. But for the time being, let us 
stress that the running backward movie picture is not self--evident. In 
particular, the interchange of initial and final state under time reversal 
is very questionable. To see this, let us first recall that there are two 
mathematical possibilities for a time reversal operator; either it must be 
unitary or anti-unitary. These lead to two quite different, both 
mathematically coherent time reversal conjugated scenarios:

The process $i\to f$ being schematized as:
\[
\left| i \right\rangle =a^+(E_{i1} )...a^+(E_{in} )\left| 0 \right\rangle \hspace{0.5cm} 
 \mbox{ }\mathop \Rightarrow \limits^{TIME ARROW} \times  \mbox{ }\hspace{0.5cm} 
\left\langle f \right|=\left\langle 0 \right|a(E_{f1} )...a(E_{fp} )
\]
\[
-\infty \leftarrow t  \hspace{5cm} t\to +\infty 
\]
the time reversed coordinate is $t_{rev} =-t$ and: 

The conventional QFT anti-unitary time reversal scenario interchanges 
initial and final states:
\[
i\to f\mbox{ }\mathop \Rightarrow \limits^{T}  \mbox{ }T^A(f)\to T^A(i)
\]
\[
\left| f \right\rangle =a^+(E_{f1} )...a^+(E_{fp} )\left| 0 \right\rangle \hspace{0.5cm} 
 \mbox{ }\mathop \Rightarrow \limits^{TIME ARROW}  \mbox{ }\hspace{0.5cm} \left\langle i \right|=\left\langle 0 \right|a(E_{i1} )...a(E_{in} )
\]
\[
-\infty \leftarrow t_{rev} \hspace{5cm} 
t_{rev} \to +\infty 
\]

The unitary one does not interchange initial and final state but reverses 
energies
\[
i\to f\mbox{ }\mathop \Rightarrow \limits^{T}  \mbox{ }T^U(i)\to T^U(f)
\]
\[
\left\langle {\tilde {f}} \right|=\left\langle 0 \right|a(-E_{f1} 
)...a(-E_{fp} ) \hspace{0.5cm} 
 \mbox{ }\mathop \Leftarrow \limits^{TIME ARROW} \mbox{ }\hspace{0.5cm} 
\left| {\tilde {i}} \right\rangle =a^+(-E_{i1} )...a^+(-E_{in} )\left| 0 
\right\rangle 
\]
\[
-\infty \leftarrow t_{rev} \hspace{5cm} 
t_{rev} \to +\infty 
\]
Our common sense intuition then tells us that the interchange of initial and 
final state, hence the anti-unitary picture stands to reason. This is 
because we naively require that in the time reverted picture the initial 
state (the ket) must come `before' the final state (the bra) i.e for a lower 
value of $t_{rev}$. However, paying careful attention to the issue we 
realize that the time arrow, an underlying concept of time flow which here 
influences our intuition is linked to a specific property of the time 
coordinate which is not relevant for a spatial coordinate, namely its 
irreversibility or causality. But as has been pointed out by many authors, 
there are many reasons to suspect that such irreversibility and time arrow 
may only be macroscopic scale (or statistical physics) valid concepts not 
making sense for a microscopic time, at least before any measurement takes 
place. We believe that our microscopic time coordinate, before measurement 
takes place, should be better considered as a spatial one, i.e possessing no 
property such as an arrow. Then, the unitary picture is the most natural one 
as a time reversal candidate process simply because it is the usual choice 
for all other discrete and continuous symmetries. 

But if neither t nor t$_{rev}$ actually stand for the genuine flowing time 
which we experiment and measure, the latter must arise at some stage and it 
is natural to postulate that its orientation corresponding to the 
experimented time arrow is simply defined in such a way that, as drawn in 
the previous pictures, the initial state (the creator) always comes before 
the final state (the annihilator) in this flowing time. This clearly points 
toward a theoretical framework where the time will be treated as a quantum 
object undergoing radical transformations from the microscopic to the 
macroscopic time we measure. Let us anticipate that the observable 
velocities will be better understood in term of this new macroscopic flowing 
time variable which arrow (orientation) keeps the same under reversal of the 
unflowing space--like t coordinate.

Therefore, the interchange of initial and final states is only justified 
under the assumption that time coordinate reversal implies time arrow 
reversal. But this is not at all obvious and thus there is no more strong 
reason to prefer and adopt the QFT anti-unitary choice. At the contrary, we 
can now list several strong arguments in favor of the unitary choice: 
\begin{itemize}
\item The mathematical handling of an anti-unitary operator is less trivial and 
induces unusual complications when applied for instance to the Dirac field.

\item  The QFT choice leads to momentum reversal, a very surprising result for a 
mass-less particle, since in this case it amounts to a genuine wavelength 
reversal and not frequency reversal, as one would have expected. 

\item Its anti-unitarity makes T really exceptional in QFT. As a consequence, not 
all basic four-vectors transform the same way under such operator as the 
reference space-time four-vector. In our mind, a basic four-vector is an 
object involving the parameters of a one particle state such as for instance 
its energy and three momentum components. The one particle state energy is 
the time component of such an object but does not reverses as the time 
itself if T is taken anti-unitary. This pseudo-vector behavior under time 
reversal seems nonsense and leads us to prefer the unitary scenario. At the 
contrary, we can understand why (and accept that) the usual operator 
four-vectors, commonly built from the fields, behave under discrete 
transformations such as unitary parity differently than the reference 
space-time four-vector. This is simply because, as we shall see, they 
involve in a nontrivial way the parity-pseudo-scalar 3-volume. 

\item  Time irreversibility at macroscopic scale allows us to define 
unambiguously our time arrow. But, as we already noticed, the arrow of time 
at the microscopic scale or before any measurement process takes place may 
be not so well defined. The statement that the time arrow is only a 
macroscopic scale (or may be statistical physics) valid concept is not so 
innovative. We know from Quantum Mechanics that all microscopic quantum 
observables acquire their macroscopic physical status through the still 
enigmatic measurement process. Guessing that the time arrow itself only 
becomes meaningful at macroscopic scale, we could reverse our microscopic 
time coordinate t as an arrowless spatial coordinate. Reverting the time 
arrow is more problematic since this certainly raises the well known time 
reversal and causality paradoxes. But the good new is that reversing the 
time coordinate does not necessarily imply reversing the arrow of time, i.e 
interchanging initial and final state. In the unitary picture, you do not 
actually go backward in time since you just see the same succession (order) 
of events counting the t$_{rev}$ time ``\`{a} rebours'', with only the signs 
of the involved energies being affected and you need not worry anymore about 
paradoxes. Therefore, in a certain sense, the running backward movie picture 
was may be just a kind of entropy reversal picture, a confusing and 
inappropriate macroscopic scale concept which obscured our understanding of 
the time coordinate reversal and led us to believe that the anti-unitary 
scenario was obviously the correct one.

\item  Charge and charge density are invariant while current densities get 
reversed under a unitary time reversal (see section VI).

\item  Negative energy fields are natural solutions of all relativistic 
equations.

\item  The instability issue might be solved in a modified general relativistic 
model as we shall show in \cite{fhc2}.
\end{itemize}
\section{ Negative energy quantum fields, time reversal and vacuum 
energies }

We shall now explicitly build the QFT neglected solutions, e.g. the usual 
bosonic and fermionic negative energy fields, show how these are linked to 
the positive ones through time reversal and how vacuum divergences cancel 
from the Hamiltonians.

\subsection{ The neutral scalar field }

The positive energy scalar field solution of the Klein-Gordon equation is:
\[
\phi (x,t)=\int {\frac{d^3p}{(2\pi )^{3/2}(2E)^{1/2}}} \left[ 
{a(p,E)e^{i(Et-px)}+a^\dag (p,E)e^{-i(Et-px)}} \right]
\]
with $E=\sqrt[]{p^2+m^2}$.
The negative energy scalar field solution of the same Klein-Gordon equation 
is:
\[
\tilde {\phi }(x,t)=\int {\frac{d^3p}{(2\pi )^{3/2}(2E)^{1/2}}} \left[ 
{\tilde {a}^\dag (-p,-E)e^{i(Et-px)}+\tilde 
{a}(-p,-E)e^{-i(Et-px)}} \right]
\]
We just required this field to create and annihilate negative energy quanta. 
Assuming T is anti-unitary, it is well known that a scalar field is 
transformed according
\[
T\phi (x,t)T^{-1}=\phi (x,-t)
\]
where, for simplicity, an arbitrary phase factor was chosen unity. Then it 
is straightforward to show that:
\[
Ta^\dag (p,E)T^{-1}=\,a^\dag (-p,E)
\]
We do not accept this result because we want time reversal to flip energy, 
not momentum. If instead, the T operator is chosen unitary like all other 
discrete transformation operators (P, C) in Quantum Field Theory we cannot 
require $T\phi (x,t)T^{-1}=\phi (x,-t)$, but rather:
\[
T\phi (x,t)T^{-1}=\tilde {\phi }(x,-t)
\]
The expected result is then obtained as usual through the change in the 
variable p$\rightarrow $-p:
\[
Ta^\dag (p,E)T^{-1}=\,\tilde {a}^\dag (p,-E)
\]
This confirms that a unitary T leads to energy reversal of scalar field 
quanta. Momentum is invariant. For a massive particle this may be 
interpreted as mass reversal coming along with velocity reversal. But in the 
unitary time reversal scenario it is not at all obvious that the velocity is 
built out of the time coordinate which gets reversed. Instead, as soon as 
this velocity is measured it seems more natural to build it out of the (as 
well measured) flowing time which never gets reversed. In this case, neither 
velocity nor mass get reversed.
The Hamiltonian for our free neutral scalar field reads:
\[
H=+\frac{1}{2}\int {d^3} x\,[\,(\frac{\partial \phi (x,t)}{\partial 
t})^2+(\frac{\partial \phi (x,t)}{\partial x})^2+m^2\phi ^2(x,t)]
\]
The Hamiltonian for the corresponding negative energy field is:
\[
\tilde {H}=\tilde {P}^0=-\frac{1}{2}\int {d^3} x\,[\,(\frac{\partial \tilde 
{\phi }(x,t)}{\partial t})^2+(\frac{\partial \tilde {\phi }(x,t)}{\partial 
x})^2+m^2\tilde {\phi }^2(x,t)]
\]
The origin of the minus sign under time reversal of $H$ will be investigated in 
sections VI. After replacing the scalar fields by their 
expressions, the computation then follows the same line as in all QFT books, 
leading to:
\[
 H=\frac{1}{2}\int {d^3} p\,p^0(a^\dag (p,E)a(p,E)+a(p,E)a^\dag 
(p,E)) \\ 
\]
\[
 \quad \tilde {H}=\;-\frac{1}{2}\int {d^3} p\,p^0(\tilde {a}^\dag 
(-p,-E)\tilde {a}(-p,-E)+\tilde {a}(-p,-E)\tilde {a}^\dag (-p,-E)) 
\\ 
\]
With $p^0=\sqrt[]{p^2+m^2}$ and the usual commutation relations, 
\[
[a_p^\dag ,a_{{p}'} ]=\delta ^4(p-{p}'),
[\tilde {a}_p^\dag ,\tilde {a}_{{p}'} ]=\delta ^4(p-{p}')\,
\]
vacuum divergences cancel (as we shall see, in a general relativistic 
framework, these only cancel as gravitational sources), and for the total 
Hamiltonian we get:
\[
H_{total} =\int {d^3} p\,p^0\left\{ {a^\dag (p,E)a(p,E)-\tilde 
{a}^\dag (-p,-E)\tilde {a}(-p,-E)} \right\}
\]
It is straightforward to check that the energy eigenvalue for a positive 
(resp negative) energy ket is positive (resp negative), as it should. For a 
vector field, the infinities would cancel in the same way assuming as well 
the usual commutation relations.
\subsection{ The Dirac field}

Let us investigate the more involved case of the Dirac field. The Dirac 
field is solution of the free equation of motion:
\[
(i\gamma ^\mu \partial _\mu -m)\psi (x,t)=0
\]
When multiplying this Dirac equation by the unitary T operator from the 
left, we get:
\[
(iT\gamma ^\mu T^{-1}\partial _\mu -TmT^{-1})T\psi (x,t)=(iT\gamma ^\mu 
T^{-1}\partial _\mu -TmT^{-1})\tilde {\psi }(x,-t)=0
\]
If the rest energy term m is related to the Higgs field value at its minimum 
(or another dynamical field) its transformation under time reversal is more 
involved than that of a pure number. Rather, we have:
\[
m=g\phi _0 (x,t)\to \tilde {m}=TmT^{-1}=g\tilde {\phi }_0 (x,-t)
\]
Making the replacement, $\partial _0 =-\partial ^0,\partial _i =\partial ^i$ 
and requiring that the T conjugated Dirac and scalar fields at its minimum 
$\tilde {\psi }(x,-t)=T\,\psi (x,t)T^{-1}$, $\,\tilde {\phi }_0 (x,-t)=T\,\phi 
_0 (x,t)T^{-1}$ together should obey the same equation, e.g. 
\[
(i\gamma ^\mu 
\partial ^\mu -g\tilde {\phi }_0 (x,-t))\tilde {\psi }(x,-t)=0
\]
 as $\psi (x,t)$ and $\phi _0 (x,t)$, leads to:
\[
T\gamma ^iT^{-1}=\gamma ^i,\,T\gamma ^0T^{-1}=-\gamma ^0
\]
The T operator is then determined to be $T=\gamma ^1\gamma ^2\gamma ^3$. Now 
assuming also that $\tilde {\phi }_0 (x,t)=-\phi _0 (x,t)$, the Dirac 
equation satisfied by $\tilde {\psi }(x,t)$ reads:
\[
(i\gamma ^\mu \partial _\mu +m)\tilde {\psi }(x,t)=0
\]
$\gamma ^0,\,\gamma ^i$ being a particular gamma matrices representation used 
in equation $(i\gamma ^\mu \partial _\mu -m)\psi (x,t)=0$, then $(i\gamma 
^\mu \partial _\mu +m)\tilde {\psi }(x,t)=0$ can simply be obtained from the 
latter by switching to the new gamma matrices representation $-\gamma 
^0,\,-\gamma ^i$ and the negative energy Dirac field $\tilde {\psi }(x,t)$. 
As is well known, all gamma matrices representations are unitary equivalent 
and here $\gamma ^5$ is the unitary matrix transforming the set $\gamma 
^0,\,\gamma ^i$ into $-\gamma ^0,\,-\gamma ^i$ ($\gamma ^5\gamma ^\mu \left( 
{\gamma ^5} \right)^{-1}=-\gamma ^\mu )$. Thus $\tilde {\psi }(x,t)$ 
satisfies the same Dirac equation as $\gamma ^5\psi (x,t)$. The physical 
consequences will be now clarified. Let us write down the positive (resp 
negative) energy Dirac field solutions of their respective equations.
\[
\psi (x,t)=\frac{1}{(2\pi )^{3/2}}\sum\limits_{\sigma =\pm 1/2} 
{\int\limits_p {\frac{d^3p}{(2E)^{1/2}}\{
u(-E,m,-p,-\sigma )a_c 
(E,m,p,\sigma )} e^{i(Et-px)}}
\]
\[
+u(E,m,p,\sigma )a^\dag (E,m,p,\sigma )e^{-i(Et-px)} \}
\]
\[
\tilde {\psi }(x,t)=\frac{1}{(2\pi )^{3/2}}\sum\limits_{\sigma =\pm 1/2} 
{\int\limits_p {\,\frac{d^3p}{(2E)^{1/2}}\{u(-E,-m,-p,-\sigma )\tilde 
{a}^\dag (-E,-m,-p,-\sigma )} e^{i(Et-px)}}
\]
\[
+u(E,-m,p,\sigma )\tilde {a}_c (-E,-m,-p,-\sigma )e^{-i(Et-px)} \} 
\]
with $E=\sqrt[]{p^2+m^2}$.
Classifying the free Dirac waves propagating in the x direction, we have as 
usual for the positive energy field spinors:
\[
\begin{array}{l}
 u(E,m,p_x ,+1/2)=\left[ {{\begin{array}{*{20}c}
 1 \hfill \\
 0 \hfill \\
 {\frac{\sigma _x p_x }{m+E}} \hfill \\
 0 \hfill \\
\end{array} }} \right],\,u(-E,m,-p_x ,-1/2)=\left[ {{\begin{array}{*{20}c}
 1 \hfill \\
 0 \hfill \\
 {\frac{\sigma _x p_x }{m-E}} \hfill \\
 0 \hfill \\
\end{array} }} \right] \\ 
 u(E,m,p_x ,-1/2)=\left[ {{\begin{array}{*{20}c}
 0 \hfill \\
 1 \hfill \\
 0 \hfill \\
 {\frac{\sigma _x p_x }{m+E}} \hfill \\
\end{array} }} \right],\,u(-E,m,-p_x ,+1/2)=\left[ {{\begin{array}{*{20}c}
 0 \hfill \\
 1 \hfill \\
 0 \hfill \\
 {\frac{\sigma _x p_x }{m-E}} \hfill \\
\end{array} }} \right] \\ 
 \end{array}
\]
The negative energy field spinors are also easily obtained through the 
replacement m $\rightarrow $ -m

$\begin{array}{l}
 u(-E,-m,-p_x ,-1/2)=\left[ {{\begin{array}{*{20}c}
 1 \hfill \\
 0 \hfill \\
 {\frac{\sigma _x p_x }{-m-E}} \hfill \\
 0 \hfill \\
\end{array} }} \right],\,u(E,-m,p_x ,1/2)=\left[ {{\begin{array}{*{20}c}
 1 \hfill \\
 0 \hfill \\
 {\frac{\sigma _x p_x }{-m+E}} \hfill \\
 0 \hfill \\
\end{array} }} \right] \\ 
 u(-E,-m,-p_x ,+1/2)=\left[ {{\begin{array}{*{20}c}
 0 \hfill \\
 1 \hfill \\
 0 \hfill \\
 {\frac{\sigma _x p_x }{-m-E}} \hfill \\
\end{array} }} \right],\,u(E,-m,p_x ,-1/2)=\left[ {{\begin{array}{*{20}c}
 0 \hfill \\
 1 \hfill \\
 0 \hfill \\
 {\frac{\sigma _x p_x }{-m+E}} \hfill \\
\end{array} }} \right] \\ 
 \end{array}$ \newline
We demand that:
\[
T\psi (x,t)T^{-1}=\tilde {\psi }(x,-t)
\]
This implies:
\[
Ta^\dag (E,m,p,\sigma )T^{-1}u(E,m,p,\sigma )=u(-E,-m,-p\to 
p,-\sigma )\tilde {a}^\dag (-E,-m,p,-\sigma )
\]
Hence:
\[
Ta^\dag (E,m,p,\sigma )T^{-1}=\tilde {a}^\dag 
(-E,-m,p,-\sigma )
\]
Thus, upon time reversal, energy, rest energy and spin are reversed. Because 
momentum is invariant helicity also flips its sign. Without having reverted 
the rest energy term in the negative energy Dirac field equation we could 
not have obtained this simple link through time reversal between the 
positive and negative energy creation operators. The rest energy reversal in 
the spinor expressions also reveals the difference between a true negative 
energy spinor $u(-E,-m,.,.)$ and a negative frequency spinor $u(-E,m,.,.)$.
\label{subsec:mylabel1}
The Hamiltonian for $\psi (x,t)$ is:
\[
H=P^0=\int {d^3} x[\bar {\psi }(x,t)(-i\gamma ^i.\partial _i +m)\psi 
(x,t)]\,+\,h.c
\]
The negative energy field Hamiltonian will be built out of negative energy 
fields explicitly different from those entering in $H$. Hence, it is hopeless 
trying to obtain such kind of simple transformation relations such as 
$P^0\Rightarrow \pm P^0$. On the other hand we can build the negative energy 
Hamiltonian and check that it provides the correct answer when applied to a 
given negative energy ket.
We know that $T\gamma ^iT^{-1}=\gamma ^i,\,T\gamma ^0T^{-1}=-\gamma ^0$,~so 
that: 
\[
T\bar {\psi }(x,t)T^{-1}=T\psi ^\dag  (x,t)\gamma ^0T^{-1}=-T\psi 
^\dag  (x,t)T^{-1}\gamma ^0
\]
\[
=-(T\psi (x,t)T^{-1})^\dag  
\gamma ^0=-\bar {\tilde {\psi }}(x,-t)
\]
This will produce an extra minus sign in the negative energy Dirac field 
Hamiltonian. The origin of the other minus sign is the same as for the 
scalar field Hamiltonian and will be clarified later. The Hamiltonian for 
$\tilde {\psi }(x,t)$ is then:
\[
\tilde {H}=\tilde {P}^0=--\int {d^3} x[\bar {\tilde {\psi }}(x,t)(-i\gamma 
^i\partial _i -m)\tilde {\psi }(x,t)]\,+\,h.c\,
\]
Because the positive (resp negative) energy spinor satisfies $(i\gamma ^\mu 
\partial _\mu -m)\psi (x,t)=0$, (resp $(i\gamma ^\mu \partial _\mu +m)\tilde 
{\psi }(x,t)=0)$ we have $(-i\gamma ^i\partial _i +m)\psi (x,t)=i\gamma 
^0\partial _0 \psi (x,t)$, (resp $(-i\gamma ^i\partial _i -m)\tilde {\psi 
}(x,t)=i\gamma ^0\partial _0 \tilde {\psi }(x,t))$. 
The Hamiltonians then read:
\[
H=P^0=i\int {d^3} x[\psi ^\dag (x,t)\partial _0 \psi 
(x,t)]\,+\,h.c\,
\]
\[
\tilde {H}=\tilde {P}^0=i\int {d^3} x[\tilde {\psi }^\dag 
(x,t)\partial _0 \tilde {\psi }(x,t)]\,+\,h.c\,
\]
Assuming for simplicity that we are dealing with a neutral field, the 
computation proceeds as usual for the positive energy Hamiltonian. With 
$p^0=\sqrt[]{p^2+m^2}$:
\[
H=\frac{1}{2}\sum\limits_{\sigma =\pm 1/2} \int {d^3} p\,p^0(a^\dag 
(E,p,\sigma )a(E,p,\sigma )-a(E,p,\sigma )a^\dag (E,p,\sigma ))
\]
Negative energy spinors possessing the same orthogonality properties as 
positive energy spinors, the negative energy Hamiltonian is then obtained by 
the simple replacements $a^\dag (E,p,\sigma )\to \tilde 
{a}(-E,-p,-\sigma )$; $a(E,p,\sigma )\to \tilde {a}^\dag 
(-E,-p,-\sigma )$:
\[
\tilde {H}=\frac{1}{2}\sum\limits_{\sigma =\pm 1/2} - \int {d^3} 
p\,p^0(\tilde {a}^\dag (-E,-p,-\sigma )\tilde {a}(-E,-p,-\sigma 
)
\]
\[
-\tilde {a}(-E,-p,-\sigma )\tilde {a}^\dag (-E,-p,-\sigma ))
\]
Infinities cancel as for the boson fields when we apply the fermionic 
anti-commutation relations $\left\{ {a_{p,\sigma }^\dag 
,a_{{p}',{\sigma }'} } \right\}=\delta ^4(p-{p}')\,\delta _{\sigma ,{\sigma 
}'} $ , $\left\{ {\tilde {a}_{p,\sigma }^\dag  ,\tilde 
{a}_{{p}',{\sigma }'} } \right\}=\delta ^4(p-{p}')\,\delta _{\sigma ,{\sigma 
}'} $, leading to:
\[
H_{total} =\sum\limits_{\sigma =\pm 1/2} \int {d^3} p\,p^0\left\{ 
{a^\dag (p,E,\sigma )a(p,E,\sigma )
-\tilde {a}^\dag 
(-p,-E,-\sigma )\tilde {a}(-p,-E,-\sigma )} \right\}
\]
It is also easily checked that the energy eigenvalue for a positive (resp 
negative) energy ket is positive (resp negative), as it should. When we 
realize how straightforward are the cancellation of vacuum divergences for 
all fields it is very tempting to state that such infinities appeared only 
because half of the field solutions were neglected! We shall show in \cite{fhc2}
that actually, in a general relativity context, our vacuum divergences only 
vanish as a source for gravitation. But the Casimir effect should still 
survive. 
\section{Phenomenology of the uncoupled positive and negative 
energy worlds}

We shall now show that the uncoupled positive and negative energy worlds are 
both perfectly viable: no stability issue arises and in both worlds the 
behavior of matter and radiation is completely similar so that the negative 
signs may just appear as a matter of convention \cite{Lin1} \cite{Lin2}. 
Consider a gas made with negative energy matter particles (fermions) and 
negative energy photons. The interaction between two negative energy 
fermions is going on through negative energy photons exchange. Because the 
main result will only depend on the bosonic nature of the considered 
interaction field, let us compute and compare the simpler propagator of the 
positive and negative energy scalar fields.

-For a positive energy scalar field:
\[
\phi (x)=\int {\frac{d^3p}{(2\pi )^{3/2}(2p^0)^{1/2}}} \left[ 
{a(p)e^{ipx}+a_c ^\dag (p)e^{-ipx}} \right]
\]
we get~as usual:
\[
 \left\langle 0 \right|T(\phi (x)\phi ^\dag (y))\left| 0 
\right\rangle =\left\langle 0 \right|\phi (x)\phi ^\dag (y)\left| 0 
\right\rangle \,\theta (x_0 -y_0 )+\left\langle 0 \right|\phi ^\dag 
(y)\phi (x)\left| 0 \right\rangle \,\theta (y_0 -x_0 )
\]
\[
 =\left\langle 0 \right|\int {\frac{d^3p}{(2\pi )^32p^0}} a(p)a^\dag 
(p)e^{ip(x-y)}\left| 0 \right\rangle \theta (x_0 -y_0 )
\]
\[
+\left\langle 0 
\right|\int {\frac{d^3p}{(2\pi )^32p^0}} a_c (p)a_c ^\dag 
(p)e^{-ip(x-y)}\left| 0 \right\rangle \theta (y_0 -x_0 )  
\]
\[
 =\int {\frac{d^3p}{(2\pi )^32p^0}} e^{ip(x-y)}\theta (x_0 -y_0 )+\int 
{\frac{d^3p}{(2\pi )^32p^0}} e^{-ip(x-y)}\theta (y_0 -x_0 ) 
\]
\[
 =\Delta (y-x)\theta (x_0 -y_0 )+\Delta (x-y)\theta (y_0 -x_0 ) \\ 
\]

-For a negative energy scalar field:
\[
\tilde {\phi }(x)=\int {\frac{d^3p}{(2\pi )^{3/2}(2p^0)^{1/2}}} \left[ 
{\tilde {a}^\dag  (p)e^{ipx}+\tilde {a}_c (p)e^{-ipx}} \right]
\]
we obtain:
\[
 \left\langle 0 \right|T(\tilde {\phi }(x)\tilde {\phi }^\dag 
(y))\left| 0 \right\rangle =\left\langle 0 \right|\tilde {\phi }(x)\tilde 
{\phi }^\dag (y)\left| 0 \right\rangle \,\theta (x_0 -y_0 
)+\left\langle 0 \right|\tilde {\phi }^\dag (y)\tilde {\phi 
}(x)\left| 0 \right\rangle \,\theta (y_0 -x_0 ) 
\]
\[
 =\left\langle 0 \right|\int {\frac{d^3p}{(2\pi )^32p^0}} \tilde {a}_c 
(p)\tilde {a}_c ^\dag (p)e^{-ip(x-y)}\left| 0 \right\rangle \theta 
(x_0 -y_0 )
\]
\[
+\left\langle 0 \right|\int {\frac{d^3p}{(2\pi )^32p^0}} \tilde 
{a}(p)\tilde {a}^\dag (p)e^{ip(x-y)}\left| 0 \right\rangle \theta 
(y_0 -x_0 ) 
 \]
\[
 =\int {\frac{d^3p}{(2\pi )^32p^0}} e^{-ip(x-y)}\theta (x_0 -y_0 )+\int 
{\frac{d^3p}{(2\pi )^32p^0}} e^{ip(x-y)}\theta (y_0 -x_0 ) 
\]
\[ 
 =\Delta (x-y)\theta (x_0 -y_0 )+\Delta (y-x)\,\theta (y_0 -x_0 )\, 
\]
Summing the two propagators, the theta functions cancel:
\[
 \left\langle 0 \right|T(\tilde {\phi }(x)\tilde {\phi }^\dag 
(y))\left| 0 \right\rangle +\left\langle 0 \right|T(\phi (x)\phi 
^\dag (y))\left| 0 \right\rangle 
\]
\[
=(\Delta (x-y)+\Delta (y-x))(\theta 
(x_0 -y_0 )+\theta (y_0 -x_0 )\,) 
\]
\[ 
 =\Delta (x-y)+\Delta (y-x)\propto \int {(\delta \left( {E-p^0} 
\right)+} \delta \left( {E+p^0} \right))e^{-iE(x_0 -y_0 )}dE \\ 
\]
Therefore, if the two propagators could contribute with the same coupling to 
the interaction between two currents, the virtual particle terms would 
cancel each other. Only on-shell particles could still be exchanged between 
the two currents provided energy momentum conservation does not forbid it. 
For a photon field as well the two off-shell parts of the propagators would 
be found opposite. Hence the coulomb potential derived from the negative 
energy photon field propagator would be exactly opposite to the coulomb 
potential derived from the positive energy photon field propagator: as a 
consequence, the 1/r Coulomb potential and electromagnetic interactions 
would simply disappear. The interesting point is that in our negative energy 
gas, where we assume that only the exchange of negative energy virtual 
photons takes place, the coulomb potential is reversed compared to the usual 
coulomb potential generated by positive energy virtual photons exchange. 
However in this repulsive potential between oppositely charged fermions, 
these still attract each other, as in the positive energy world, because of 
their negative inertial terms in the equation of motion (as deduced from 
their negative terms in the action). The equation of motion for a given 
negative energy matter particle in this Coulomb potential is:
\[
-m\dot {v}=--\frac{\partial U_c }{\partial r}
\]
or
\[
m\dot {v}=-\frac{\partial U_c }{\partial r}
\]
We find ourselves in the same situation as that of a positive energy 
particles gas interacting in the usual way e.g through positive energy 
photons exchange. Hence negative energy atoms will form and the main results 
of statistical physics apply: following Boltzman law, our particles will 
occupy with the greatest probabilities states with minimum $\frac{1}{2}m\dot 
{v}^2$, thus with maximum energy$-\frac{1}{2}m\dot {v}^2$. Temperatures are 
negative. This result can be extended to all interactions propagated by 
bosons as are all known interactions. The conclusion is that the non-coupled 
positive and negative energy worlds are perfectly stable, with positive and 
negative energy particles minimizing the absolute value of their energies:~
\section{ Actions and Hamiltonians under Time reversal and Parity}

\subsection{ Negative integration volumes?}

Starting from the expression of the Hamiltonian density for a positive 
energy neutral scalar field:
\[
T^{00}(x,t)=\left( {\frac{\partial \phi (x,t)}{\partial t}} 
\right)^2+\sum\limits_{i=1,3} {\left( {\frac{\partial \phi (x,t)}{\partial 
x_i }} \right)^2+m^2\phi ^2(x,t)} 
\]
and applying time reversal we get: 
\[
\left( {\frac{\partial \tilde {\phi }(x,-t)}{\partial t}} 
\right)^2+\sum\limits_{i=1,3} {\left( {\frac{\partial \tilde {\phi 
}(x,-t)}{\partial x_i }} \right)^2+m^2\tilde {\phi }^2(x,-t)} 
\]
with $T\phi (x,t)T^{-1}\equiv \tilde {\phi }(x,-t)$
From such expression, a naive free Hamiltonian density for the scalar 
field$\tilde {\phi }(x,t)$ may be proposed: 
\[
\tilde {T}^{00}(x,t)=\left( {\frac{\partial \tilde {\phi }(x,t)}{\partial 
t}} \right)^2+\sum\limits_{i=1,3} {\left( {\frac{\partial \tilde {\phi 
}(x,t)}{\partial x_i }} \right)^2+m^2\tilde {\phi }^2(x,t)} 
\]
It thus happens that $\tilde {T}^{00}(x,t)$ is manifestly positive since it 
is a sum of squared terms. We of course cannot accommodate negative energy 
fields with positive Hamiltonian densities so following the procedure used 
to obtain negative kinetic energy terms for a phantom field, we just assumed 
in the previous sections a minus sign in front of this expression. But how 
could we justify this trick if time reversal does not provide us with this 
desired minus sign? One possible solution appears when we realize that 
according to general relativity, actually $T^{00}$ is not a spatial energy 
density but rather$\sqrt g T^{00}$ where $g\equiv -Det\,g_{\mu \nu } $. This 
is also expected to still remain positive because of a rather strange 
mathematical choice in general relativity: integration volumes such as $dt$, 
$d^4x$, $d^3x$ are not signed and should not flip sign under time reversal 
or parity transformations. 
Let us try the more natural opposite way: $t\to -t{\kern 1pt}{\kern 
1pt}\Rightarrow dt\to -dt{\kern 1pt} $ and $x\to -x{\kern 
1pt}{\kern 1pt}\Rightarrow dx\to -dx$, natural in the sense that this is 
naively the straightforward mathematical way to proceed and let us 
audaciously imagine that for instance a negative 3-dimentional volume is 
nothing else but the image of a 3-dimentional positive volume in a mirror. 
Then, the direct consequence of working with signed volumes is that the 
general relativistic integration element $d^4x\sqrt g $ is not invariant 
anymore under coordinate transformations (such as P or T) with negative 
Jacobian (it is often stated that the absolute value of the Jacobian is 
imposed by a fundamental theorem of integral calculus[2]. But should not 
this apply only to change of variables and not general coordinate 
transformations?). We are then led to choose an invariant integration 
element~under any coordinate transformations: this is $d^4x\left| 
{\frac{\partial \xi }{\partial x}} \right|$, where $\left| {\frac{\partial 
\xi }{\partial x}} \right|$ stands for the Jacobian of the transformation 
from the inertial coordinate system $\xi ^\alpha$ to $x^\mu $. Because 
$\left| {\frac{\partial \xi }{\partial x}} \right|$ is not necessarily 
positive as is $\sqrt g$ in general relativity, it will get reversed under P 
or T transformations affecting Lorentz indices only so that spatial charge 
density $\left| {\frac{\partial \xi }{\partial x}} \right|J^0$, scalar 
charge $Q=\int {\left| {\frac{\partial \xi }{\partial x}} \right|J^0} d^3x$, 
spatial energy-momentum densities $\left| {\frac{\partial \xi }{\partial x}} 
\right|T^{\mu 0}$ and energy-momentum four-vector $P^\mu =\int {\left| 
{\frac{\partial \xi }{\partial x}} \right|T^{\mu 0}} d^3x$ should transform 
accordingly. For instance, it is often stated that a unitary time reversal 
operator is not allowed because it would produce the not acceptable charge 
reversal. This analysis is no more valid if the Jacobi determinant flips its 
sign. Indeed, though $J^0$, as all four-vector time components, becomes 
negative, the spatial charge density $\left| {\frac{\partial \xi }{\partial 
x}} \right|J^0$ and scalar charge $Q=\int {\left| {\frac{\partial \xi 
}{\partial x}} \right|J^0} d^3x$ remain positive under unitary time 
reversal.
It is also worth checking what is now the effect of unitary space inversion:
$P^\mu =\int {\left| {\frac{\partial \xi }{\partial x}} \right|T^{\mu 0}} 
d^3x$ transforms under Parity as $T^{\mu 0}$ times the pseudo-scalar Jacobi 
determinant $\left| {\frac{\partial \xi }{\partial x}} \right|$, so that:
\[
P^0\Rightarrow -P^0,\,P^i\Rightarrow P^i
\]
$Q=\int {\left| {\frac{\partial \xi }{\partial x}} \right|J^0} d^3x$ also 
transforms under Parity as $J^0$ times the pseudo-scalar Jacobi determinant 
$\left| {\frac{\partial \xi }{\partial x}} \right|$, so that:
\[
Q\Rightarrow -Q
\]
So, if unitary Parity has the same effect on various fields, currents and 
energy densities as in conventional quantum field theory, it now produces a 
flip in the energy and charge signs but does not affect momentum! 
Anyway, we see that the signed Jacobi determinant could do the good job for 
providing us with the desired minus signs. However, working with negative 
integration volumes amounts to give up the usual definition of the integral 
which insures that it is positive definite. If we are not willing to give up 
this definition, another mechanism should be found to provide us with the 
necessary minus sign. The issue will be reexamined and a more satisfactory 
solution described in \cite{fhc2}.

\section{ Interactions between positive and negative energy fields ?}

Postulate the existence of a new inertial coordinate system $\tilde {\xi 
}$ such that $\left| {\frac{\partial \tilde {\xi }}{\partial x}} \right|$ is 
negative. This can be achieved simply by considering the two time reversal 
conjugated (with opposite proper times) inertial coordinate systems $\xi 
$ and $\tilde {\xi }$. We may then define the positive energy quantum $F(x)$ 
fields (resp negative energy $\tilde {F}(x)$ fields) as the fields entering in 
the action with positive $\left| {\frac{\partial \xi }{\partial x}} \right|$ 
(resp negative $\left| {\frac{\partial \tilde {\xi }}{\partial x}} \right|)$ 
entering in the integration volume so that the energy $P^0=\int {\left| 
{\frac{\partial \xi }{\partial x}} \right|T^{00}} d^3x$ (resp $\tilde 
{P}^0=\int {\left| {\frac{\partial \tilde {\xi }}{\partial x}} \right|\tilde 
{T}^{00}} d^3x)$ is positive (resp negative). The action for positive energy 
matter and radiation is then as usual:
\[
S=\int {d^4} x\left| {\frac{\partial \xi }{\partial x}} \right|\left\{ 
{L(\Psi (x),\frac{\partial \xi ^\alpha }{\partial x^\mu }(x))+L(A_\mu 
(x),\frac{\partial \xi ^\alpha }{\partial x^\mu }(x))+J_\mu (x)A^\mu (x)} 
\right\}
\]
Similarly, the action for negative energy matter and radiation is:
\[
\tilde {S}=\int {d^4} x\left| {\frac{\partial \tilde {\xi }}{\partial x}} 
\right|\left\{ {L(\tilde {\Psi }(x),\frac{\partial \tilde {\xi }^\alpha 
}{\partial x^\mu }(x))+L(\tilde {A}_\mu (x),\frac{\partial \tilde {\xi 
}^\alpha }{\partial x^\mu }(x))+\tilde {J}_\mu (x)\tilde {A}^\mu (x)} 
\right\}
\]
Hence positive energy fields move under the influence of the gravitational 
field $\frac{\partial \xi ^\alpha }{\partial x^\mu }$, while negative energy 
fields move under the influence of the gravitational field $\frac{\partial 
\tilde {\xi }^\alpha }{\partial x^\mu }$. Then, the mixed coupling in the 
form $J_\mu (x)\tilde {A}^\mu (x)$ that we might have naively hoped is not 
possible just because the integration volume must be $d^4x\left| 
{\frac{\partial \xi }{\partial x}} \right|$ for $F(x)$ type fields and 
$d^4x\left| {\frac{\partial \tilde {\xi }}{\partial x}} \right|$ for $\tilde 
{F}(x)$ type fields. Indeed coherence requires that in the action the 
negative Jacobian be associated with negative energy fields $\tilde {F}(x)$ 
involving negative energy quanta creation and annihilation operators. This 
is a good new since it is well known that couplings between positive and 
negative energy fields lead to an unavoidable stability problem due to the 
fact that energy conservation keeps open an infinite phase space for the 
decay of positive energy particles into positive and negative energy 
particles. A scenario with positive and negative energy fields living in 
different metrics also provides a good way to account for the undiscovered 
negative energy states. 
However the two metrics should not be independent if we want to introduce a 
connection at least gravitational between positive and negative energy 
worlds, mandatory to make our divergences gravitational effects actually 
cancel. In \cite{fhc2} we shall explicit this dependency between the two conjugated 
metrics and the mechanism that gives rise through the extremum action 
principle to the negative source terms in the Einstein equation. It will be 
clear that this mechanism only works properly if, as in general relativity, 
we keep working with Jacobi determinants absolute values and do not give up 
the usual definition of integrals.

\section{ Maximal C, P and baryonic asymmetries }

One of the most painful concerns in High Energy Physics is related to our 
seemingly inability to provide a satisfactory explanation for the maximal 
Parity violation observed in the weak interactions. The most popular model 
that may well account, through the seesaw mechanism, for the smallness of 
neutrino masses is quite disappointing from this point of view since parity 
violation is just put in by hand, as it is in the standard model, in the 
form of different spontaneous symmetry breaking scalar patterns in the left 
and right sectors. The issue is just postponed, and we are still waiting for 
a convincing explanation for this trick. Actually, one gets soon convinced 
that the difficulty comes from the fact that Parity violation apparently 
only exists in the weak interaction. Much more easy would be the task to 
search for its origin if this violation was universal. And yet, quite 
interestingly, it seems possible to extend parity violation to all 
interactions, just exploiting the fundamental structure of fermion fields 
and at the same time explain why this is only detectable and apparent in the 
weak interactions. 
There exists four basic degrees of freedom, solutions of the Dirac field 
equations: these are $\psi _L (x),\psi _R (x),\psi _{cL} (x),\psi _{cR} (x)$ 
but two of them suffice to create and annihilate quanta of both charges and 
helicities: for instance the usual $\psi (x)=\psi _L (x)+\psi _R (x)$ may be 
considered as the most general Dirac solution:
\[
\psi (x)=\frac{1}{(2\pi )^{3/2}}\int\limits_{p,\sigma } {u(p,\sigma )a_c 
(p,\sigma )} .e^{i(px)}+v(p,\sigma )a^\dag (p,\sigma 
)e^{-i(px)}\;d^3p
\]
But another satisfactory base, as far as our concern is just to build 
kinetic{\_}interaction terms and not mass terms, could be the pure left 
handed $\psi _L (x)+\psi _{cL} (x)$ field making use of the charge 
conjugated field. 
\[
\psi _c (x)=\frac{1}{(2\pi )^{3/2}}\int\limits_{p,\sigma } {u(p,\sigma 
)a(p,\sigma )\,} e^{i(px)}+v(p,\sigma )a_c^{^\dag } (p,\sigma 
)e^{-i(px)}\;d^3p
\]
Indeed, from a special relativistic mass-less Hamiltonian such as 
\[
H_L^0 =\int {d^3} x[\psi _L ^{^\dag }(-i\alpha .\nabla )\psi _L 
]\,+\int {d^3} x[\psi _{cL} ^{^\dag }(-i\alpha .\nabla )\psi _{cL} 
]\,
\]
the same normal ordered current and physics as the usual one are derived 
when requiring various global symmetries to become local (this is checked in 
the Annex). 
\[
\overline \Psi \gamma _\mu \frac{1-\gamma _5 }{2}\Psi (x)-\overline {\Psi _c 
} \gamma _\mu \frac{1-\gamma _5 }{2}\Psi _c (x)=:\overline \Psi 
\gamma _\mu \Psi (x):
\]
Assume now that the corresponding general Dirac field built out of only 
right handed components is not redundant with the previous (as is generally 
believed because except for a Majorana particle, both create and annihilate 
quanta of all charges and helicities) but lives in the conjugated metric, an 
assumption which we shall later justify. This then would be from our world 
point of view a negative energy density field. This manifestly maximal 
parity violating framework would not allow to detect any parity violating 
behavior in those interactions involving only charged Dirac particles in 
their multiplets, because the charge conjugated left handed field $\psi _{cL} (x)$
can successfully mimic the right handed field $\psi _R (x)$. However, 
in any interaction involving a completely neutral e.g Majorana fermion, 
$\psi _{cL} (x)$ could not play this role anymore resulting as in the weak 
interaction in visible maximal parity and charge violation (we claim that 
though no symmetry forbids it, the one degree of freedom Majorana field for 
a neutrino cannot be associated simultaneously with the two degrees of 
freedom of the Dirac charge field, since this amounts to duplicate the 
Majorana kinetic term and appears as an awkward manipulation, therefore one 
has to choose which electron/positron charge is associated with the neutral 
particle (neutrino) in the multiplet, this resulting in maximal charge 
violation and making the already present parity violation manifest). Even 
neutral-less fermion multiplets as in the quark sector of the weak 
interactions could then have inherited this parity and charge violation 
provided their particles lived together with neutral fermion particles in 
higher dimensional groups before symmetry breaking occurred producing their 
separation into distinct multiplets.

Now what about mass terms? For charged fields, coupling with a positive 
energy right handed field must take place to produce the chirality flipping 
mass term. But the right handed field is not there. It may be that no bare 
mass term is explicitly allowed to appear in an action and that a new 
mechanism should be found to produce interaction generated massive 
propagators starting from a completely mass-less action. Let us guess that 
such scenario is not far from the one which is actually realized in nature, 
because maximal Charge and Parity violation, and the related bayonic 
asymmetry of the universe has otherwise all of the characteristics of a not 
solvable issue.

But why should right handed chiral fields be negative energy density fields? 
Because this is what the pseudo-vector behavior under Parity of the operator 
four-momentum told us in section VI.1. Remember however that the unitary 
parity conjugated field creates positive energy point-like quanta and can be 
viewed as a positive point-like energy field (this is a standard QFT 
result). This four-vector behavior under parity of the one particle state 
four-momentum (an object we called a basic four-vector in section III.2) 
seems to be in contradiction with the pseudo-vector behavior of the 
four-momentum field operator. Actually the measured energy of the particle 
is obtained by acting with the energy operator on the one particle state 
ket. So there is no contradiction because this measurement is of course 
performed on a non zero three dimensional volume and we have to admit that 
the measured energy of the particle we get is negative from our world point 
of view as a result of the particle being living in an enantiomorphic 
3-dimensionnal space. In other words, from our world point of view, the 
parity conjugated field has a negative energy \textbf{density,} which we may 
consider as a positive energy per negative inertial 3-volume, so that it 
leads to a negative energy when integrated on a general coordinate 3-volume 
(as if it was a parity scalar, the behavior of this 3-volume under a parity 
transformation plays no role in our discussion since this is just one of the 
general coordinate transformations). 
Then the PT fields are again positive integrated energy (energy measured in 
a finite volume) fields but oppositely charged (charge measured in a finite 
volume), i.e describing anti-particles (see VI.1) living in our world metric 
and interacting with their PT symmetric fields describing particles. 

In short, we believe that recognizing the universality of Parity violation, 
i.e the fact that we are living in a left chiral world, is also an 
interesting approach to the issue. It then suffices to introduce the right 
chiral parity conjugated world (its action) to plainly restore Parity 
invariance of the total action. Eventually it may be, as already Sakharov 
suggested in 1967 \cite{Sak} , that we are living in a left chiral positive energy 
world with its particles and antiparticles while the conjugated world is 
from our world point of view a right chiral negative energy world with its 
particles and antiparticles.

\section{ Synthesis}

Let us gather the main information we learned from our investigation of 
negative energies in Relativistic QFT indicating that the correct 
theoretical framework for handling them should be found in a modified GR.

\begin{itemize}
\item The TheoreticaI Viewpoint
\end{itemize}

In second quantification, all relativistic field equations admit genuine 
negative energy field solutions creating and annihilating negative energy 
quanta. Unitary time reversal links these fields to the positive energy 
ones. The unitary choice, usual for all other symmetries in physics also 
allows to avoid the well known paradoxes associated with time reversal. 
Positive and negative energy fields vacuum divergences we encounter after 
second quantization are unsurprisingly found to be exactly opposite. The 
negative energy fields action must be maximised. However there is no way to 
reach a coherent theory involving negative energies in flat space-time. 
Indeed, if positive and negative energy scalar fields are time reversal 
conjugated, their Hamiltonian densities and actions must also be so which we 
shall find to be only possible in the context of general relativity thanks 
to the metric transformation under discrete symmetries.

\begin{itemize}
\item The Phenomenological Viewpoint
\end{itemize}

In a mirror negative energy world which fields remain non coupled to our 
world positive energy fields, stability is insured and the behavior of 
matter and radiation is as usual. Hence, it's just a matter of convention to 
define each one as a positive or negative energy world. Only if they could 
interact, would we expect hopefully promising new phenomenology since many 
outstanding enigmas, among which are the flat galactic rotation curves, the 
Pioneer effect, the universe flatness, acceleration and its voids, indicate 
that repelling gravity might play an important role in physics. On the other 
hand, negative energy states never manifested themselves up to now, strongly 
suggesting that a barrier is at work preventing the two worlds to interact 
except through gravity. 

\begin{itemize}
\item The Main Issues
\end{itemize}

A trivial cancellation between vacuum divergences is not acceptable since 
the Casimir effect shows evidence for vacuum fluctuations. But in our 
approach, the positive and negative energy worlds will be maximally 
gravitationally coupled in such a way as to only produce exact cancellations 
of vacuum energies gravitational effects. Also, a generic catastrophic 
instability issue arises whenever quantum positive and negative energy 
fields are allowed to interact. If we restrict the stability issue to our 
modified gravity we will see that this disastrous scenario is also avoided. 
At last, allowing both positive and negative energy virtual photons to 
propagate the electromagnetic interaction simply makes it disappear. The 
local gravitational interaction will be treated very differently in our 
modified GR so that this unpleasant feature also be avoided.

\begin{itemize}
\item Outlooks
\end{itemize}

A left-handed kinetic and interaction Lagrangian can satisfactorily describe 
all known physics except mass terms which anyway remain problematic in 
modern physics. This strongly supports the idea that the right handed chiral 
fields might be living in another world (where the 3-volume reversal under 
parity presumably would make these fields acquire a negative energy density) 
and may provide an interesting explanation for maximal parity violation 
observed in the weak interaction.

If the connection between the two worlds is fully reestablished above a 
given energy threshold, then loop divergences naturally would get cancelled 
thanks to the positive and negative energy virtual propagators compensation. 
Such reconnection might take place through a new transformation process 
allowing particles to jump from one metric to the conjugated one \cite{Chen} 
presumably at places where the conjugated metrics meet each other. 

\section{ Conclusion}

Of course, negative energy matter remains undiscovered at present and the 
stability issue strongly suggests that making it interact with normal matter 
requires new non standard interaction mechanisms. However, considering the 
seemingly many related theoretical and phenomenological issues and recalling 
the famous historical examples of equation solutions that were considered 
unphysical for a long time before they were eventually observed, we believe 
it is worth trying to understand how negative energy solutions should be 
handled in GR. We will propose a special treatment for discrete symmetry 
transformations in GR. A new gravitational picture will be derived in \cite{fhc2} 
opening rich phenomenological and theoretical perspectives and making us 
confident that the approach is on the right way.

\section{Acknowledgments} I above all thank my parents but also my friends 
and family for their staunch support. I specially thank J.P Petit, J. Mallet, G. 
D'agostini and P. Midy who accepted to check carefully my computations. It is a pleasure 
to warmly thank A. Tilquin, J-B. Devivie, T. Schucker, G. Esposito-Far\`{e}se and D. 
Buskulic for many helpful advices, comments and fascinating discussions. I 
acknowledge my colleagues and friends from the Centre de Physique des 
Particules de Marseille, the Centre de Physique Th\'{e}orique and the 
Facult\'{e} de Luminy for their kindness and interest in my work, 
particularly P. Coyle, E. Nagy, M. Talby, E. Kajfasz, F. Bentosela, 
J. Busto, O. Leroy, A. Duperrin and my friends from the RENOIR group.

\newpage
\section{Annex}

For a purely left-handed kinetic lagrangien,
\[
L_{kin} =-\bar {\Psi }_L \gamma ^\mu \partial_\mu \Psi _L -\bar {\Psi }_{Lc}
 \gamma ^\mu  \partial_\mu \Psi _{Lc} 
\]
Gauge invariance yields interaction terms~:
\[
L_{kin} +L_{int} =-\bar {\Psi }_L (\gamma ^\mu [\partial _\mu +ieA_\mu 
])\Psi _L -\bar {\Psi }_{Lc} (\gamma ^\mu [\partial _\mu -ieA_\mu ])\Psi _{Lc} 
\]
from which follows the QED current~:
\[
[\bar {\Psi }\gamma _\mu (\frac{1-\gamma _5 }{2})\Psi (x)-\bar {\Psi }_c 
\gamma _\mu (\frac{1-\gamma _5 }{2})\Psi _c (x)]
\]
\textbf{Useful formula (\cite{Wein2} p219{\&}225) }
\[
u^+(q,\sigma )=(u^\ast (q,\sigma ))^T=\left( {-\beta Cv(q,\sigma )} 
\right)^T=-v(q,\sigma )^T C^T \beta ^T
\Rightarrow u^+(q,\sigma )=v(q,\sigma )^TC \beta 
\]
\[
v^+(q,\sigma )=(v^\ast (q,\sigma ))^T=\left( {-\beta Cu(q,\sigma )} 
\right)^T=-u(q,\sigma )^T C^T \beta ^T
\]
\[
\Rightarrow v^+(q,\sigma )=u(q,\sigma )^T C\beta \;(1)
\]
then
\[
u^+({q}',{\sigma }')\beta \gamma _\mu \frac{1-\gamma _5 }{2}u(q,\sigma 
)=v({q}',{\sigma }')^TC\gamma _\mu \frac{1-\gamma _5 }{2}u(q,\sigma )
\]
\[
v^+(q,\sigma )\beta \gamma _\mu \frac{1+\gamma _5 }{2}v({q}',{\sigma 
}')=u(q,\sigma )^TC\gamma _\mu \frac{1+\gamma _5 }{2}v({q}',{\sigma }')
\]
using
\[
\begin{array}{l}
 (C\gamma _\mu \frac{1-\gamma _5 }{2})^T=\frac{1-\gamma _5 ^T}{2}\gamma _\mu 
^TC^T=-\frac{1-\gamma _5 ^T}{2}\gamma _\mu ^TC \\ 
 =\frac{1-\gamma _5 ^T}{2}C\gamma _\mu =C\frac{1-\gamma _5 }{2}\gamma _\mu 
=C\gamma _\mu \frac{1+\gamma _5 }{2} \\ 
 \end{array}
\]
we obtain the first useful formula
\[
u^+({q}',{\sigma }')\beta \gamma _\mu \frac{1-\gamma _5 }{2}u(q,\sigma 
)=v^+(q,\sigma )\beta \gamma _\mu \frac{1+\gamma _5 }{2}v({q}',{\sigma }')
\]
From (1) we get
\[
\begin{array}{l}
 v^+(q,\sigma )\beta \gamma _\mu \frac{1-\gamma _5 }{2}u({q}',{\sigma 
}')=u(q,\sigma )^TC\gamma _\mu \frac{1-\gamma _5 }{2}u({q}',{\sigma }') \\ 
 \quad \quad \quad \quad =u({q}',{\sigma }')^TC\gamma _\mu \frac{1+\gamma _5 
}{2}u(q,\sigma ) \\ 
 \end{array}
\]
but
\[
v^+({q}',{\sigma }')\beta \,=u({q}',{\sigma }')^TC
\]
which leads to the second useful formula
\[
v^+(q,\sigma )\beta \gamma _\mu \frac{1-\gamma _5 }{2}u({q}',{\sigma 
}')=v^+({q}',{\sigma }')\beta \,\gamma _\mu \frac{1+\gamma _5 }{2}u(q,\sigma 
)
\]
\newpage
\textbf{Computation of the left-handed current }
\[
 \overline \Psi \gamma _\mu \frac{1-\gamma _5 }{2}\Psi (x)= 
\]
\[ 
 \frac{1}{(2\pi )^3}\int\limits_{p,{p}',\sigma ,{\sigma }'} {u^\ast 
({p}',{\sigma }')\beta \gamma _\mu \,} \frac{1-\gamma _5 }{2}u(p,\sigma 
).e^{i(-{p}'x+px)}a^\dag ({p}',{\sigma }')a(p,\sigma )d^3pd^3{p}' \\ 
 \]
\[
 +\frac{1}{(2\pi )^3}\int\limits_{p,{p}',\sigma ,{\sigma 
}'} {v^\ast ({p}',{\sigma }')\beta \gamma _\mu \,\frac{1-\gamma _5 }{2}} 
v(p,\sigma ).e^{i({p}'x-px)}a_c ({p}',{\sigma }')a_c^\dag  (p,\sigma 
)d^3pd^3{p}'
\]
\[
+\frac{1}{(2\pi )^3}\int\limits_{p,{p}',\sigma ,{\sigma 
}'} {v^\ast ({p}',{\sigma }')\beta \gamma _\mu \frac{1-\gamma _5 }{2}\,} 
u(p,\sigma ).e^{i({p}'x+px)}a_c ({p}',{\sigma }')a(p,\sigma )d^3pd^3{p}'
\]
\[
 +\frac{1}{(2\pi )^3}\int\limits_{p,{p}',\sigma ,{\sigma 
}'} {u^\ast ({p}',{\sigma }')\beta \gamma _\mu \frac{1-\gamma _5 }{2}\,} 
v(p,\sigma ).e^{i(-{p}'x-px)}a^\dag ({p}',{\sigma }')a_c^\dag  
(p,\sigma )d^3pd^3{p}'
\]
\[
 \overline {\Psi _c } \gamma _\mu \frac{1-\gamma _5 }{2}\Psi _c (x)= 
\]
\[ 
 \frac{1}{(2\pi )^3}\int\limits_{p,{p}',\sigma ,{\sigma }'} {v^\ast 
(p,\sigma )\beta \gamma _\mu \frac{1-\gamma _5 }{2}\,} v({p}',{\sigma 
}').e^{i(px-{p}'x)}a(p,\sigma )a^\dag  ({p}',{\sigma }')d^3pd^3{p}' 
\]
\[
\frac{1}{(2\pi )^3}\int\limits_{p,{p}',\sigma ,{\sigma }'} {u^\ast (p,\sigma 
)\beta \gamma _\mu \frac{1-\gamma _5 }{2}\,} u({p}',{\sigma 
}').e^{i(-px+{p}'x)}a_c ^\dag (p,\sigma )a_c ({p}',{\sigma 
}')d^3pd^3{p}'
\]
\[
\frac{1}{(2\pi )^3}\int\limits_{p,{p}',\sigma ,{\sigma }'} {u^\ast 
(p,\sigma )\beta \gamma _\mu \frac{1-\gamma _5 }{2}\,} v({p}',{\sigma 
}').e^{i(-px-{p}'x)}a_c ^\dag (p,\sigma )a^\dag  
({p}',{\sigma }')d^3pd^3{p}'
\]
\[
\frac{1}{(2\pi )^3}\int\limits_{p,{p}',\sigma ,{\sigma }'} {v^\ast (p,\sigma 
)\beta \gamma _\mu \frac{1-\gamma _5 }{2}\,} u({p}',{\sigma 
}').e^{i(px+{p}'x)}a(p,\sigma )a_c ({p}',{\sigma }')d^3pd^3{p}'
\]
\[
 \overline {\Psi _c } \gamma _\mu \frac{1-\gamma _5 }{2}\Psi _c (x)=
\]
\[ 
 \frac{1}{(2\pi )^3}\int\limits_{p,{p}',\sigma ,{\sigma }'} {u^\ast 
({p}',{\sigma }')\beta \gamma _\mu \frac{1+\gamma _5 }{2}\,} u(p,\sigma 
).e^{i(px-{p}'x)}a(p,\sigma )a^\dag ({p}',{\sigma }')d^3pd^3{p}' \\ 
\]
\[
+\frac{1}{(2\pi )^3}\int\limits_{p,{p}',\sigma ,{\sigma }'} {v^\ast 
({p}',{\sigma }')\beta \gamma _\mu \frac{1+\gamma _5 }{2}\,} v(p,\sigma 
).e^{i(-px+{p}'x)}a_c ^\dag (p,\sigma )a_c ({p}',{\sigma 
}')d^3pd^3{p}'
\]
\[
+\frac{1}{(2\pi )^3}\int\limits_{p,{p}',\sigma ,{\sigma }'} {u^\ast 
({p}',{\sigma }')\beta \gamma _\mu \frac{1+\gamma _5 }{2}\,} v(p,\sigma 
).e^{i(-px-{p}'x)}a_c ^\dag (p,\sigma )a^\dag  ({p}',{\sigma 
}')d^3pd^3{p}'
\]
\[
+\frac{1}{(2\pi )^3}\int\limits_{p,{p}',\sigma ,{\sigma }'} {v^\ast 
({p}',{\sigma }')\beta \gamma _\mu \frac{1+\gamma _5 }{2}\,} u(p,\sigma 
).e^{i(px+{p}'x)}a(p,\sigma )a_c ({p}',{\sigma }')d^3pd^3{p}'
\]
\[
\begin{array}{l}
 \overline {\Psi _c } \gamma _\mu \frac{1-\gamma _5 }{2}\Psi _c (x)= \\ 
 -\frac{1}{(2\pi )^3}\int\limits_{p,{p}',\sigma ,{\sigma }'} {u^\ast 
({p}',{\sigma }')\beta \gamma _\mu \frac{1+\gamma _5 }{2}\,} u(p,\sigma 
).e^{i(px-{p}'x)}a^\dag  ({p}',{\sigma }')a(p,\sigma )d^3pd^3{p}' \\ 
 \end{array}
\]
\[
-\frac{1}{(2\pi )^3}\int\limits_{p,{p}',\sigma ,{\sigma }'} {v^\ast 
({p}',{\sigma }')\beta \gamma _\mu \frac{1+\gamma _5 }{2}\,} v(p,\sigma 
).e^{i(-px+{p}'x)}a_c ({p}',{\sigma }')a_c ^\dag (p,\sigma 
)d^3pd^3{p}'
\]
\[
-\frac{1}{(2\pi )^3}\int\limits_{p,{p}',\sigma ,{\sigma }'} {u^\ast 
({p}',{\sigma }')\beta \gamma _\mu \frac{1+\gamma _5 }{2}\,} v(p,\sigma 
).e^{i(-px-{p}'x)}a^\dag  ({p}',{\sigma }')a_c ^\dag 
(p,\sigma )d^3pd^3{p}'
\]
\[
-\frac{1}{(2\pi )^3}\int\limits_{p,{p}',\sigma ,{\sigma }'} {v^\ast 
({p}',{\sigma }')\beta \gamma _\mu \frac{1+\gamma _5 }{2}\,} u(p,\sigma 
).e^{i(px+{p}'x)}a_c ({p}',{\sigma }')a(p,\sigma )d^3pd^3{p}'
\]
\[
+\frac{1}{(2\pi )^3}\int\limits_{p,\sigma } {u^\ast (p,\sigma )\beta \gamma 
_\mu \frac{1+\gamma _5 }{2}\,} u(p,\sigma ).d^3p
\]
\[
+\frac{1}{(2\pi )^3}\int\limits_{p,\sigma } {v^\ast (p,\sigma )\beta \gamma 
_\mu \frac{1+\gamma _5 }{2}\,} v(p,\sigma ).d^3p
\]
\[
 \overline \Psi \gamma _\mu \frac{1-\gamma _5 }{2}\Psi (x)-\overline {\Psi 
_c } \gamma _\mu \frac{1-\gamma _5 }{2}\Psi _c (x)= 
\]
\[ 
 \frac{1}{(2\pi )^3}\int\limits_{p,{p}',\sigma ,{\sigma }'} {u^\ast 
({p}',{\sigma }')\beta \gamma _\mu \,} u(p,\sigma 
).e^{i(px-{p}'x)}a^\dag  ({p}',{\sigma }')a(p,\sigma )d^3pd^3{p}' \\ 
\]
\[
\frac{1}{(2\pi )^3}\int\limits_{p,{p}',\sigma ,{\sigma }'} {v^\ast 
({p}',{\sigma }')\beta \gamma _\mu \,} v(p,\sigma ).e^{i(-px+{p}'x)}a_c 
({p}',{\sigma }')a_c ^\dag (p,\sigma )d^3pd^3{p}'
\]
\[
\frac{1}{(2\pi )^3}\int\limits_{p,{p}',\sigma ,{\sigma }'} {u^\ast 
({p}',{\sigma }')\beta \gamma _\mu \,} v(p,\sigma 
).e^{i(-px-{p}'x)}a^\dag ({p}',{\sigma }')a_c ^\dag 
(p,\sigma )d^3pd^3{p}'
\]
\[
\frac{1}{(2\pi )^3}\int\limits_{p,{p}',\sigma ,{\sigma }'} {v^\ast 
({p}',{\sigma }')\beta \gamma _\mu } u(p,\sigma ).e^{i(px+{p}'x)}a_c 
({p}',{\sigma }')a(p,\sigma )d^3pd^3{p}'
\]
\[
-\frac{1}{(2\pi )^3}\int\limits_{p,\sigma } {v^\ast (p,\sigma )\beta \gamma 
_\mu \,} v(p,\sigma ).d^3p
\]
At last:
\[
\overline \Psi \gamma _\mu \frac{1-\gamma _5 }{2}\Psi (x)-\overline {\Psi _c 
} \gamma _\mu \frac{1-\gamma _5 }{2}\Psi _c (x)=:\overline \Psi 
\gamma _\mu \Psi (x):
\]
For a Majorana field, $\Psi _c (x)$ is not there and we are left only with a 
chiral kinetic term:
\[
\overline \Psi \gamma _\mu \frac{1-\gamma _5 }{2}\Psi (x)
\]
We believe that such term cannot be duplicated to be found associated in 
multiplets with both $\Psi (x)$ and $\Psi _c (x)$ of a Dirac field, so that 
the above chiral kinetic term will necessarily result in a chiral 
interaction term in which parity and charge violation explicitly manifest 
themselves.
\end{document}